\documentclass[twocolumn]{aastex61}
\usepackage{silence}
\usepackage{graphicx}
\usepackage[caption=false]{subfig}
\usepackage{booktabs}
\usepackage{natbib}
\setcitestyle{authoryear,open={(},close={)}}

\usepackage{float}
\usepackage{color,soul}

\usepackage{wrapfig}
\usepackage{bm}
\usepackage{amsmath}

\usepackage{fancyhdr}
\usepackage{wasysym}

\newcommand{\lir}{L$_{\mathrm{\scriptsize IR}}$}

\newcommand{\fagn}{$f_{\mathrm{\scriptsize AGN, IR}}$}

\newcommand{\lagnbol}{L$_{\mathrm{AGN}}$}
\newcommand{\lsfbol}{L$_{\mathrm{SF}}$}
\newcommand{\lagnoff}{L$_{\lambda}^{\mathrm{AGN,off}}$}
\newcommand{\lagntorus}{L$_\lambda^{\mathrm{AGN,10x,input}}$}
\newcommand{\lstellar}{L$_\lambda^{\mathrm{stellar,input}}$}
\newcommand{\tpeak}{$t_{peak}$}
\newcommand{\leff}{L$_{\lambda}^{\mathrm{AGN,eff}}$}

\newcommand{\ledd}{$\lambda_{edd}$}

\newcommand{\cchb}[1]{\textcolor{black}{#1}}

\newcommand{\cch}[1]{\textcolor{black}{#1}}
\newcommand{\jm}[1]{\textcolor{black}{#1}}

\begin{document}

\title{\cchb{Dust-enshrouded} AGN can dominate host-galaxy-scale cold-dust emission}
\author{Jed McKinney} 
\affil{Department of Astronomy, University of Massachusetts, Amherst, MA 01003, USA}
\email{jhmckinney@umass.edu}
\author{Christopher C. Hayward}
\affil{Center for Computational Astrophysics, Flatiron Institute, 162 Fifth Avenue, New York, NY 10010, USA}
\author{Lee J. Rosenthal}
\affil{Department of Astronomy, California Institute of Technology, Pasadena, CA 91125, USA}
\author{Juan Rafael Mart\'inez-Galarza}
\affil{Center for Astrophysics $\vert$ Harvard \& Smithsonian, 60 Garden Street, Cambridge, MA 02138, USA}
\author{Alexandra Pope}
\affil{Department of Astronomy, University of Massachusetts, Amherst, MA 01003, USA}
\author{Anna Sajina}
\affil{Department of Physics \& Astronomy, Tufts University, Medford, MA 02155, USA}
\author{Howard A. Smith}
\affil{Center for Astrophysics $\vert$ Harvard \& Smithsonian, 60 Garden Street, Cambridge, MA 02138, USA}

\begin{abstract}
It is widely assumed that long-wavelength infrared (IR) emission from cold dust ($T\sim20-40\,$K) is a reliable tracer of star formation even in the presence of a bright active galactic nucleus (AGN). Based on radiative transfer (RT) models of clumpy AGN tori, hot dust emission from the torus contributes negligibly to the galaxy spectral energy distribution (SED) at $\lambda\ga100$ \micron. However, these models do not include AGN heating of host-galaxy-scale diffuse dust, which may have far-IR (FIR) colors comparable to cold diffuse dust heated by stars. To quantify the contribution of AGN heating to host-galaxy-scale cold dust emission at $\lambda\ga100$ \micron, we perform dust RT calculations on a simulated galaxy merger both including and excluding the bright AGN that it hosts. By differencing the SEDs yielded by RT calculations with and without AGN that are otherwise identical, we quantify the FIR cold dust emission arising solely from re-processed AGN photons. In extreme cases, AGN-heated host-galaxy-scale dust can increase galaxy-integrated FIR flux densities by factors of 2-4; star formation rates calculated from the FIR luminosity assuming no AGN contribution can overestimate the true value by comparable factors. Because the FIR colors of such systems are similar to those of purely star-forming galaxies and redder than torus models, broadband SED decomposition may be insufficient for disentangling the contributions of stars and heavily dust-enshrouded AGN in the most IR-luminous galaxies. We demonstrate how kpc-scale resolved observations can be used to identify deeply dust-enshrouded AGN with cool FIR colors when spectroscopic and/or X-ray detection methods are unavailable.

\end{abstract}

\section{Introduction} \label{sec:intro}
Most of the stellar mass in galaxies was assembled between a redshift of $z=3$ to 1 (e.g., \citealt{CarilliWalter2013,MadauDickinson2014}); at $z=2$, the star-formation rate density (SFRD) of the Universe peaked, fueled by enhanced gas accretion from the intergalactic medium \citep{Keres2005,Keres2009,Genzel2008,Tacconi2010}. Statistical analysis of X-ray and infrared (IR) observations of galaxies hosting active galactic nuclei (AGN) support the coincident mass growth of central supermassive black holes (SMBHs;  \citealt{Shankar2009,Aird2010,Delvecchio2014}) and possibly SMBH-galaxy co-evolution (e.g., \citealt{Ferrarese2000,Gebhardt2000,Kormendy2013,MadauDickinson2014}). 
In particular, AGN feedback may be responsible for quenching star formation \citep[e.g.,][]{Springel2005,Cicone2014} or/and maintaining quenched galaxies by suppressing cooling flows \citep[e.g.,][]{Keres2005,Croton2006,Bower2006}.
AGN feedback helps cosmological theories of galaxy evolution match observations -- in particular, the number counts of massive and quenched galaxies at $z=0$ (e.g., \citealt{Bower2006,Nelson2018}). 

Testing this framework against observations, particularly at high redshift, is limited by the ability to separate the AGN and star-forming components of the spectral energy distributions (SEDs) of galaxies,
as to test the `quasar mode' of AGN feedback, it is necessary to simultaneously constrain the AGN luminosity and star formation rate (SFR) at a time when the AGN should be heavily \cchb{dust-enshrouded}
\citep[e.g.,][]{DiMatteo2005,Hopkins2008}.
The far-IR (FIR; $\lambda\gtrsim100\,\mu$m)/sub-millimeter (sub-mm) regime is thus a powerful probe of dust-obscured mass assembly.
It has long been known that \cchb{dust-enshrouded} AGN can dominate the total IR luminosity, powered predominantly by warm dust emitting at $\mathrm{5\,\mu m<\lambda_{rest}\leq30\,\mu m}$  \citep[e.g.,][]{Sanders1988}, but it is
typically assumed that AGN contribute negligibly to FIR emission at wavelengths longer than rest-frame $100\,\mu$m,
as predicted by torus models \citep[e.g.,][]{Fritz2006,Nenkova2008a,Nenkova2008b}.
Consequently, cold dust emission at rest-frame wavelengths $\gtrsim100\,\mu$m is sometimes treated as a `safe'
SFR tracer even when an AGN dominates the bolometric luminosity of the galaxy by explicitly
converting the FIR luminosity into an SFR value
\citep[e.g.,][]{Hatziminaoglou2010,Kalfountzou2014,Azadi2015,Gurkan2015,Stacey2018,Banerji2018,Wethers2020}.

Alternatively, the full SED may be fit using an SED
modeling code that includes an AGN component
\citep[e.g.,][]{Ciesla2015,Chang2017,Lanzuisi2017,Dietrich2018,Leja2018,Pouliasis2020,Ramos2020,Yang2020}.
Such SED modeling codes
generally assume that AGN contribute
negligibly to cold-dust emission longward of $\sim 100$
\micron, and rely
on SED model libraries generated by performing RT
on AGN torus models. The resulting SEDs
include hot dust emission from the torus
\emph{but not potential AGN-powered cold
dust emission on host-galaxy scales by construction},
even though UV/optical photons emitted by the accretion
disk or IR photons from the torus can in principle
heat ISM dust directly and/or by progressive attenuation and re-radiation at longer wavelengths.
That this possibility is realized in
some systems is suggested by results indicating that
the host galaxy can be responsible for a
non-negligible fraction of the obscuring column density
in some AGN (\citealt{Hickox2018}
and references therein). 

The origin of the FIR emission in galaxies hosting luminous AGN has been debated over the past few decades (e.g., \citealt{Downes1998,Page2001,Franceschini2003,Ruiz2007,Kirkpatrick2012,Kirkpatrick2015}). Over the years, a growing body of literature has been making the case for AGN-powered FIR emission at wavelengths $\gtrsim100\,\mu$m in QSOs (e.g., \citealt{Sanders1989,Yun1998,Nandra2007,Petric2015,Schneider2015,Symeonidis2016,Symeonidis2017,Symeonidis2018}); however, other works find conflicting results (e.g., \citealt{Stanley2017,Shangguan2020,DiMascia2021}). Spatially resolved observations of heavily obscured nearby systems find that a significant fraction of the galaxy-integrated FIR/sub-mm emission  can originate from very small regions (e.g., $\lesssim15\,$pc for Arp 220; \citealt{Scoville2017}), suggestive that deeply dust-obscured AGN power this emission (as it is unlikely that a star-forming region forming stars at a rate $\ga 100\,\mathrm{M_\odot\,yr^{-1}}$ would be this compact). However, the fact that these nuclear regions can be opaque well into the FIR makes it difficult to 
conclusively identify the power source, and how to infer the amount of
host-galaxy, kpc-scale dust heating due to AGN emission is unclear.

\jm{A promising method for decomposing galaxy-integrated FIR SEDs into AGN and SF components directly relies on mid-IR spectroscopy (e.g., \citealt{Pope2008}). \cite{Kirkpatrick2012,Kirkpatrick2015} use this technique to subtract AGN-heated dust from the SEDs of a sample of high$-z$ dusty galaxies using mid-IR spectroscopic methods assuming all of the $\sim80-90$ K dust emission is attributed to the AGN alone. These authors find that up to 75\% of \lir\ can be powered by AGN heating of the hot dust in galaxies with spectroscopic signatures of AGN in the mid-IR. Whether this conclusion extends to host-galaxy-scale cold dust emission at wavelengths past $100\,\mu$m is difficult to test empirically because of sparse wavelength coverage in the rest-frame FIR/sub-mm and because dust re-processing erases information about the original energy source. \citet{Roebuck2016} applied \citeauthor{Kirkpatrick2015}'s method to attempt to recover the AGN fraction of simulated galaxies, including the one employed in this work. Although the recovered and true AGN fractions were in qualitative agreement, \cite{Roebuck2016} found that SED decomposition underestimated the AGN contribution to the IR luminosity in some cases.}

In this work, we \cch{build on previous work in which we used hydrodynamical simulations plus radiative transfer (RT) calculations to investigate the effects of galaxy-scale dust reprocessing of AGN torus emission \citep{Younger2009,Snyder2013,Roebuck2016}.
Specifically, we use a hydrodynamical simulation of an equal-mass galaxy merger post-processed with RT to} investigate the possibility that heating from a luminous AGN embedded within a dusty galaxy can power host-galaxy-scale cold dust emission at rest-frame wavelengths $\lambda \ga 100\,\mu$m, well beyond the shorter-wavelength regime in which significant AGN powering of dust continuum flux densities is well accepted by the community \citep{Sanders1988,Casey2014}.
We create initial conditions for two identical
galaxies on a parabolic orbit.
The hydrodynamical simulation evolves the system,
including models for star formation, stellar feedback,
black hole accretion and AGN feedback.
Then, we perform dust RT calculations in
post-processing for various time snapshots to
compute spatially resolved UV-mm SEDs of the simulated galaxy merger.
By performing dust RT calculations both including and excluding the emission from the AGN, we can explicitly identify IR emission powered by the AGN on both torus and host-galaxy (kpc) scales.
\cch{This particular simulation is a massive, gas-rich,
merger designed to be analogous to $z \sim 2-3$
submillimeter galaxies (SMGs), and it is indeed
consistent with many properties of observed SMGs
\citep{Hayward2011,Hayward2012,Hayward2013}.
We chose this particular simulation because
it exhibits a high SFR, a high AGN luminosity, and high
dust attenuation; such systems are the most likely
candidates for AGN powering host-galaxy scale cold-dust
emission.}
\cch{We show that in this admittedly extreme system,}
\cchb{dust-enshrouded} AGN can be the dominant power source of cold-dust emission on host-galaxy scales, with cool FIR colors indistinguishable from those of
purely star-forming galaxies. We estimate the potential bias in observations of galaxy SFRs 
and discuss implications for SED decomposition.

The remainder of this paper is organized as follows: in Section \ref{sec:methods}, we discuss the details of our numerical simulation and RT calculations.
Section \ref{sec:results} explains how we extract the AGN-powered dust emission and presents our main results.
We discuss the implications of these calculations for inferred trends in galaxy evolution and limitations of this work in Section \ref{sec:discussion}. We summarize in Section \ref{sec:conclusions}. Throughout this work, we assume a Salpeter IMF and adopt a $\Lambda$CDM cosmology with $\Omega_m=0.3$, $\Omega_\Lambda=0.7$, and $H_0=70$ km s$^{-1}$ Mpc$^{-1}$. 

\section{Simulation and Radiative Transfer Details\label{sec:methods}}

This work makes use of the results of RT calculations originally presented in \citet{Snyder2013}.
Our re-analysis focuses on the output from a \texttt{Gadget-2} simulation \citep{Springel2005} of an idealized (non-cosmological) major merger of two identical disk galaxies.
The initial conditions were generated following \citet{Springel2005merger}.
The initial halo and baryonic masses are $9\times10^{12}$ M$_\odot$ and $4\times10^{11}$ M$_\odot$, respectively.
The initial black hole mass is $1.4\times10^5$ M$_\odot$, and the initial gas fraction is 60\%. Star formation and stellar feedback are modeled as described in \citet{Springel2003} and \citet{Springel2005merger}.
Black holes grow via Eddington-limited Bondi-Hoyle accretion. 

The RT code \texttt{SUNRISE} \citep{Jonsson2006,Jonsson2010} was used to compute SEDs for seven viewing angles every 10 Myr
throughout the simulation run. 
\cchb{The outputs of the hydrodynamical simulation are used as input for the RT calculations, i.e., the former is used to specify the 3D distribution of sources of emission (stars and AGN) and dust. 
The ages and metallicities of the star particles
are used to assign single-age stellar population SEDs to individual star particles.
The metal distribution determines the dust density distribution; a dust-to-metals ratio of 0.4 was assumed.} For more in-depth discussion of the simulation assumptions, sub-grid models, and numerical methods, see \cite{Snyder2013} and \cite{Hayward2011}. We stress that while the sub-grid models employed in this work are admittedly limited approximations of reality, they have been tested in previous works and are found to yield galaxies with properties broadly consistent with high$-z$ dust-obscured galaxies. With this in mind, this study should be considered a somewhat idealized numerical experiment designed to cleanly identify reprocessed dust emission powered by AGN in a `reasonable' simulated merger rather than an attempt to perfectly reproduce reality.

For each time snapshot of the merger simulation, different RT calculations were performed;
they differ only in how the luminosity of the AGN is computed. The `fiducial' RT runs use the accretion rate from the simulation to compute the
AGN luminosity, assuming 10\% radiative efficiency.
\cchb{We use luminosity-dependent AGN SED templates derived from observations of unreddened QSOs \citep{Hopkins2007c} as the input SEDs emitted by the AGN particle(s)}; see \citet{Snyder2013} for details. 
In the `AGN-off' runs, the luminosity of the AGN is artificially set to zero (i.e., only stellar emission is considered in the radiative
transfer calculations). We emphasize that AGN feedback is still included because the same hydrodynamical simulation is used as input, so these runs determine the impact of the AGN emission on the total SED \emph{all else being equal}, including feedback from the AGN on the host-galaxy ISM. Finally, in the `AGN-10x' RT calculations, \cch{the luminosity of the AGN is artificially boosted by a factor of 10
(equivalent to assuming a radiative efficiency of 100\%
or an instantaneous accretion rate ten times the value
computed in the simulation). As in the `AGN-off'
runs, the AGN feedback
is not altered (the same hydrodynamical simulations
are used) because we wish to isolate the effect of
the AGN on the SED}. In other words, the input of a boosted AGN spectrum into the RT calculations is the only difference between the AGN-10x and AGN-off results. Notably, this technique ignores how increasing the AGN luminosity by a factor of 10 would affect the evolution of the system;
this is \emph{by design}, as to extract the AGN-powered IR emission, we need all other aspects of the run (e.g., the dust geometry and SFR) to be fixed.

\cite{Snyder2013} performed these calculations simply to span a larger range in the AGN fractional contribution to the bolometric luminosity,
but these should not be considered `unphysical' by any means, as the relatively crude accretion model employed and the limited resolution of the hydrodynamical simulation
may cause the BH accretion rate to be underestimated \citep[e.g.,][]{Hayward2014b,Angles-Alcazar2020}. 
The Eddington ratio (\ledd$\equiv \mathrm{L}_{\mathrm{AGN}}/\mathrm{L}_{edd}$) varies from zero to one in the fiducial runs and thus spans a range of \ledd$\,=0-10$ in the AGN-10x runs.
There is both theoretical \citep[e.g.,][]{Begelman2002,Jiang2014} and observational \citep[e.g.,][]{Kelly2013,Shirakata2019} evidence for super-Eddington accretion.

\begin{figure}[t!]
	\centering
	\includegraphics[width=0.48\textwidth]{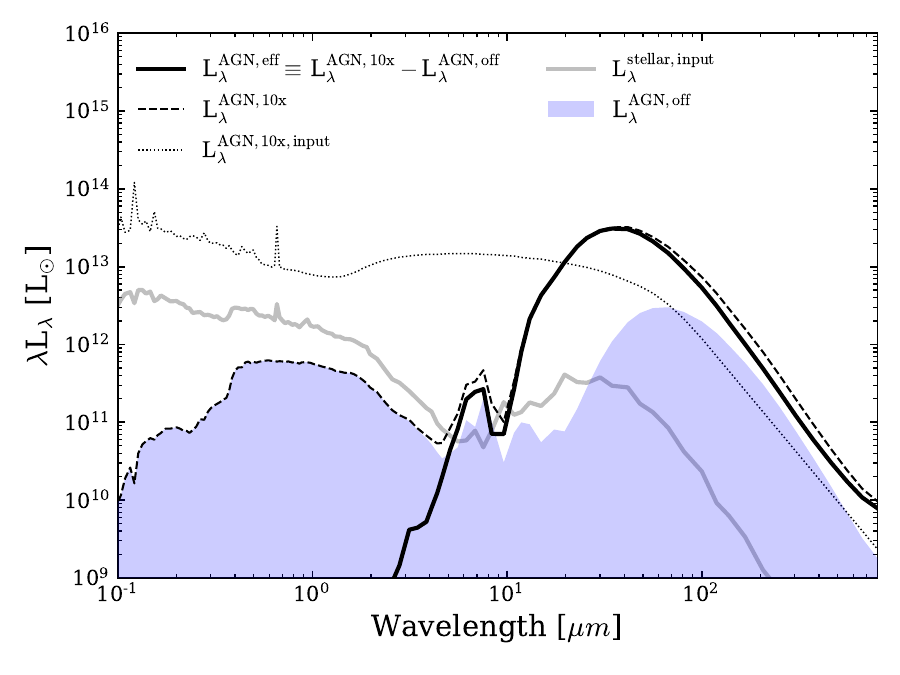}
	\caption{\footnotesize Rest-frame SED for the AGN-10x model (dashed line) at the time of peak \lagnbol/\lsfbol, where the AGN dominates the bolometric luminosity and the time at which
	the AGN contributes maximally to the FIR luminosity during this simulation. The unnatenuated AGN torus spectrum (\lagntorus) is shown as a dotted line, and the unnatenuated stellar spectrum is shown as a solid grey line (\lstellar). For the same simulation
	snapshot, we show the corresponding AGN-off SED shaded in blue. The effective AGN SED, calculated by taking the difference between the AGN-10x and AGN-off SEDs, is
	indicated by the solid line. This corresponds to the primary torus emission attenuated by dust \emph{plus} host-galaxy-scale thermal dust emission powered by the AGN.
	Due to heavy dust obscuration experienced by the AGN, the AGN contributes negligibly to the UV-optical SED, but it dominates longward of a
	few microns, including the FIR cold dust emission traditionally associated with star formation. 
	\label{fig:sed}}
\end{figure}

\begin{figure*}
    \centering
    \includegraphics[width=0.95\textwidth]{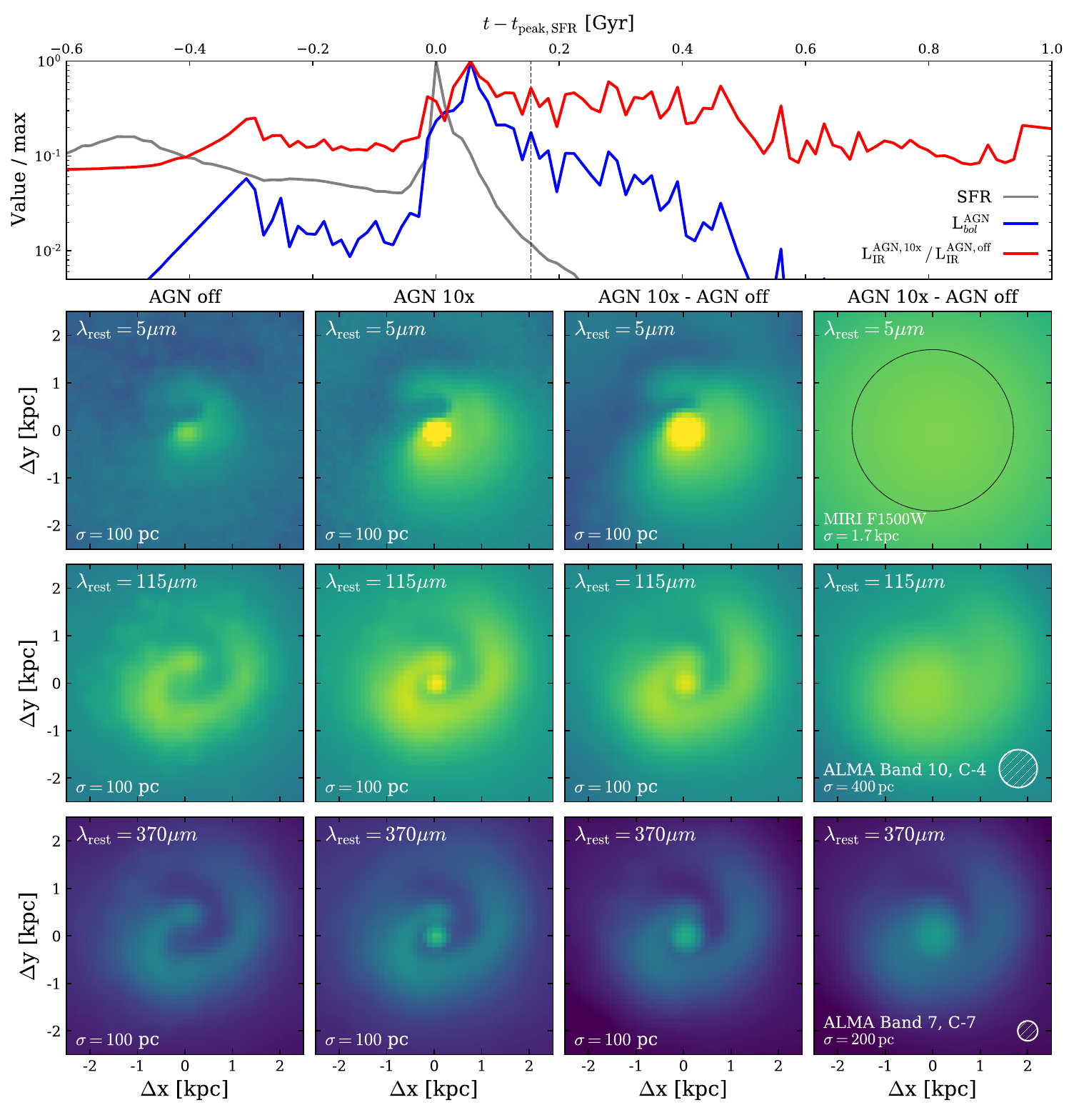}
    \caption{Emission maps of the simulated merger at a representative time (marked by the dashed line in the top panel) following the peak SFR and peak bolometric AGN luminosity. The first (second) column
    shows maps of the emission at rest-frame wavelengths $15\,\mu$m, $115\,\mu$m, and $370\,\mu$m for the AGN-off (AGN-10x) RT run. The third column shows difference images
    between the AGN-10x and AGN-off maps, i.e., the dust emission powered directly by the AGN. The color scale and stretch is constant across all panels.
    For comparison with observations, we blur the images to $100$ pc resolution, typical of spatially resolved studies of lensed systems at $z\sim2$. In the far-right column, we \jm{further blur each map to the spatial resolution achievable with \textit{JWST}/MIRI (\textit{top}) and ALMA high-frequency observations in extended array configurations} (\textit{center, bottom})  for a galaxy at $z=2$. \jm{The \textit{JWST}/MIRI beam is shown as a solid black circle, and the ALMA beams as white hatched circles.} The AGN-powered cold dust emission includes both a compact nuclear
    component and an extended component spanning a few kpc.
    \label{fig:extent}}
\end{figure*}

\section{Results\label{sec:results}}

For every snapshot and each of the seven viewing angles, we calculate the SED of the AGN-powered dust emission, i.e., the dust emission implicit in the input torus model SED
plus the host-galaxy dust emission powered by the AGN (rather than stars). Our \texttt{SUNRISE} calculations with and without AGN emission are otherwise identical; therefore, taking the difference
between the SEDs with and without AGN emission yields the SED of all photons of AGN origin, including those reprocessed by dust. Originally demonstrated in \cite{Roebuck2016} for our fiducial simulation
(see their Fig. 2),
this differencing technique is shown in Figure \ref{fig:sed} for the AGN-10x run at \tpeak, the simulation time when the ratio of bolometric AGN luminosity (\lagnbol) to bolometric stellar luminosity
(\lsfbol) reaches its maximal value of $\sim10$. At this time, the mass of the BH is $3.2\times10^8\,\mathrm{M_\odot}$, and it is accreting at a rate of $82\,\mathrm{M_\odot\,yr^{-1}}$, while SFR$\,=470\,\mathrm{M_\odot\,yr^{-1}}$. The SEDs shown in Fig.~\ref{fig:sed} are the following: 
\begin{itemize}
    \item $\mathrm{L_\lambda^{AGN,\,10x}}$\,-- The attenuated$+$re-radiated SED corresponding to the AGN-10x run, where the luminosity of the AGN is artifically boosted by a factor of 10.
    \item $\mathrm{L_\lambda^{AGN,\,off}}$\,-- The attenuated$+$re-radiated SED corresponding to the RT run where the luminosity of the AGN is set to zero.
    \item \leff\,-- The ``effective AGN SED'' calculated from the difference between $\mathrm{L_\lambda^{AGN,\,10x}}$ and $\mathrm{L_\lambda^{AGN,\,off}}$, which removes stellar-heated dust emission, leaving behind only AGN photons, \cchb{including IR photons from the torus that are absorbed by dust in the ISM and} re-radiated at progressively longer wavelengths. 
    \item \lagntorus\,-- The input AGN SED (from \citealt{Hopkins2007c}) in the AGN-10x run. \cchb{This SED is integrated from 0.1--1000 \micron} to calculate \lagnbol. 
    \item $\mathrm{L_\lambda^{stellar,\,input}}$\,-- The unattenuated stellar spectrum, which is the same across all RT runs. \cchb{This SED is integrated over 0.1--1000 \micron} to determine \lsfbol.
\end{itemize}
\leff\ peaks at $31\,\mu$m, corresponding to an effective dust temperature (from Wien's law) of $96$ K, on the warm end of dust temperatures observed in AGN-host galaxies \citep{Kirkpatrick2015}.
\jm{In the simulated SEDs,} most of the IR emission between 10 and 50 $\mu$m comes from re-processed AGN photons, consistent with numerous observations of local and high$-z$ ultraluminous infrared galaxies (ULIRGs; \lir$>10^{12}$ L$_\odot$) hosting AGN \citep[e.g.,][]{Genzel1998,Sajina2007}. Notably, \leff\ is greater than \lagnoff\ by a factor of $\gtrsim2$ from $100-1000\,\mu$m. In this particular extreme case, even the coldest dust is predominantly heated by the AGN.

\cchb{The top panel of Fig.~\ref{fig:extent} shows
the time evolution of the SFR, AGN luminosity, and fractional contribution of the AGN to the IR luminosity. Near coalescence of the two galaxies,
there is a strong starburst. The SFR then rapidly decays due to gas consumption (there is no cosmological gas accretion in this idealized simulation) and AGN feedback. Approximately 50
Myr after the peak SFR, the AGN luminosity peaks.
During this period, both the newly formed stars 
and AGN are deeply obscured by dust.
The AGN contributes $\ga 30\%$ of the IR luminosity
for $\sim 0.5$ Gyr, starting at the peak of the
starburst. The AGN luminosity then declines, decreasing below
10\% of its peak luminosity $\sim 100$ Myr later.}

The lower part of Fig.~\ref{fig:extent} presents maps of the \jm{5\,$\mu$m, 115\,$\mu$m and 370\,$\mu$m} dust emission in the AGN-off (first column) and AGN-10x (second column) runs $\sim150$ Myr after \tpeak, 
\jm{a time at which there is significant dust-obscured star formation but the \cchb{dust-enshrouded} AGN dominates the luminosity (\lagnbol/\lsfbol$>1$); this time is marked by
the dashed vertical line in the top panel.
Given the short duration of the starburst and
peak in AGN luminosity, most observations of
heavily dust-obscured AGN should probe this
phase (subsequent to the peak SFR and AGN luminosity),
so these maps should be considered `representative'.
Fig.~\ref{fig:extent} also shows} the difference between the \jm{AGN-10x and AGN-off simulations} (third column), which \jm{captures} the spatial extent of the dust emission \emph{powered by the AGN}. These maps have been convolved with a Gaussian
with FWHM of 100 pc, \jm{representative} of spatially resolved studies of lensed systems at $z\sim2-3$ \jm{(e.g., \citealt{Swinbank2015,Sharda2018,Canameras2017,Massardi2018})}.

\begin{figure}
    \centering
    \includegraphics[width=0.48\textwidth]{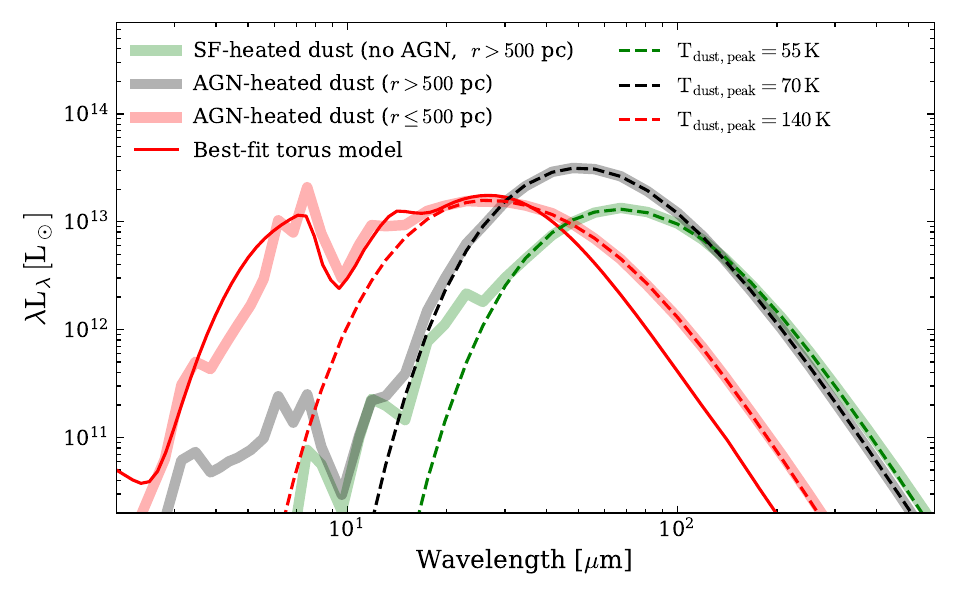}
    \caption{Rest-frame SED of AGN-heated dust emission including direct  heating and heating from dust self-absorption in the central kpc of the simulation (red, $r\leq500$ pc) and in the extranuclear region (black, $r>500$ pc), both extracted from the difference map (i.e., Fig.\,\ref{fig:extent}, col.\,3), $\sim150$ Myr after \tpeak. The green SED corresponds to diffuse star formation-heated dust at $r>500$ pc in the AGN-off run (Fig.\,\ref{fig:extent}, col.\,1). Dashed lines show modified blackbody models that best fit the $r>500$ pc SEDs at $\lambda>10\,\mu$m. The solid red line shows the best fit dusty, AGN torus model from the \cite{Fritz2006} library to the AGN-heated dust within the central kpc, which exhibits $\mathrm{T_{dust}=140}$ K from Wien's law. The AGN-heated dust emission within the central kpc is dominated by the hot, dusty torus, whereas the diffuse extended component is well described by a modified blackbody with $\mathrm{T_{dust}=70}$ K, warmer than the star formation-heated diffuse dust, but cooler than the torus emission.
    }\label{fig:regionSED}
\end{figure}

The third column of Fig.\,\ref{fig:extent} shows that the cold dust emission at \jm{FIR wavelengths $\gtrsim100\,\mu$m} powered by the AGN consists of a compact nuclear component and a lower-surface-brightness
extended component on kpc scales, clearly demonstrating that photons from the AGN heat cold dust throughout the host galaxy on kpc scales. Indeed, the integrated emission within the central kpc is well fit by a hot, dusty torus model from the \cite{Fritz2006} library, as shown in Figure \ref{fig:regionSED}. On the other hand, the host-galaxy-scale AGN-heated diffuse dust is well fit by a modified blackbody with $\mathrm{T_{dust}=70}$ K, warmer than the kpc-scale star formation-heated dust by 15 K, but significantly cooler than the dusty torus emission within the central kpc. Notably, 22\% (78\%) of the reprocessed AGN luminosity between $8-1000$ \micron\ ($100-500$ \micron) arises from outside the central kpc. The innermost region powers $\sim25$\% of \lir\ (accounting for star formation and AGN heating) integrated across the map. Comparable fractions of \lir\ arising from the centers of galaxies have been measured in nearby systems with spatially resolved FIR observations such as in the west nucleus of Arp 220 \citep{Scoville2017}.

\begin{figure}
    \centering
    \includegraphics[width=0.45\textwidth]{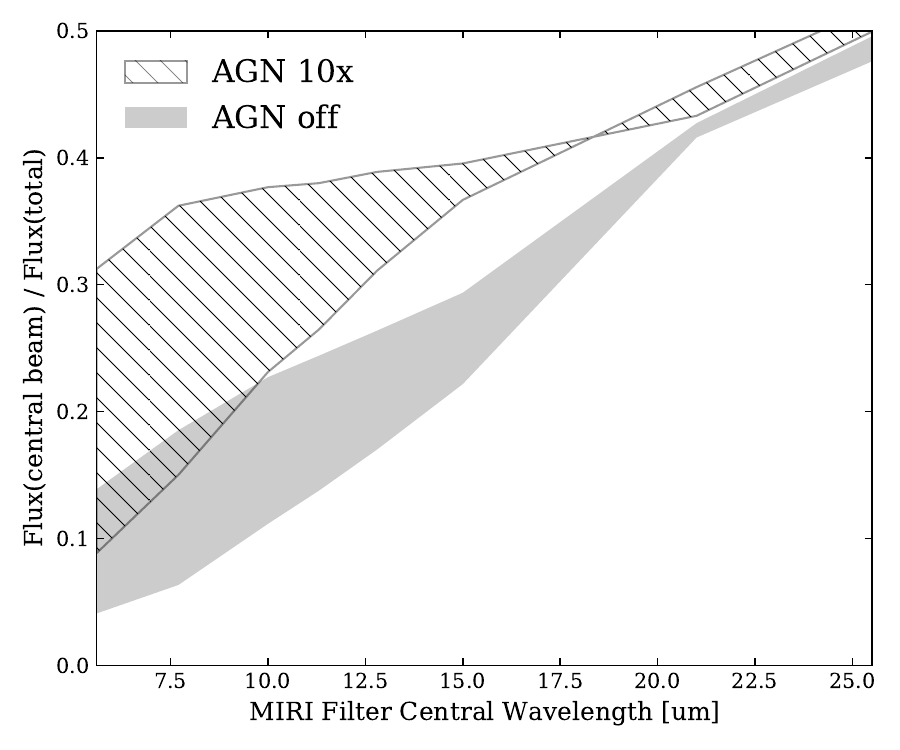}
    \caption{The ratio of flux within a simulated \textit{JWST}/MIRI beam placed on the center of the galaxy relative to the total flux across the PSF-convolved map, for various MIRI filters. We show the results of such mock observations for the AGN-10x (hatched) and AGN-off (solid gray) runs at \tpeak, the same snapshot used in Fig. \ref{fig:extent}. The maps have been redshifted to $z=2$ and convolved with the wavelength-dependent PSF of \textit{JWST}/MIRI. The shaded and hatched regions contain the range of simulated observations for all seven viewing angles. The presence of a powerful, dust-enshrouded AGN boosts the fraction of the total flux within the central beam by a factor of $\sim1.5$ between observed wavelengths of $\sim6-15\,\mu$m. Mid-IR imaging at \textit{JWST}'s resolution could be used to identify such \cchb{dust-enshrouded} AGN.}
    \label{fig:cntr_flux}
\end{figure}

\jm{In the fourth column of Figure \ref{fig:extent}, we} show maps of the AGN-powered dust emission convolved to the spatial resolution that \textit{JWST}/MIRI (\textit{top}) and ALMA's high-frequency, extended configurations (\textit{center, bottom}) can achieve for an un-lensed galaxy at $z=2$. The nuclear dust heated by the \cchb{dust-enshrouded} AGN can be resolved from the extended cold dust with ALMA.
At coarser spatial resolutions, imaging with \textit{JWST}/MIRI can test for buried AGN heating host-galaxy dust by comparing central and extended emission on scales within and outside $\sim2$ kpc (solid circle), nearly twice the 1.1 kpc half-mass radius of the system. In Figure \ref{fig:cntr_flux}, we show the ratio of the flux within a simulated \textit{JWST}/MIRI beam placed on the center of a mock observation to the total flux in the map, for MIRI filters with central wavelengths between 5 and 25$\,\mu$m. Before the simulated aperture photometry, we first convolve the maps to the spatial resolution of JWST assuming the galaxy is at $z=2$. The presence of the luminous \cchb{dust-enshrouded} AGN increases the fraction of the total flux within the central region relative to the total flux by a factor of $\sim1.5$ in most MIRI filters. As expected, the AGN-10x simulation exhibits more concentrated emission in all MIRI filters than the AGN-off run because of the strong, nuclear dust-heating source. 
IR imaging and spatially resolved SED fitting will be key for identifying such \cchb{dust-enshrouded} AGN  because, as we shall see below, the cold FIR colors would preclude distinguishing this source from a compact nuclear starburst based on the galaxy-integrated broadband FIR SED alone.

\begin{figure*}
	\centering
	\includegraphics[width=.9\textwidth]{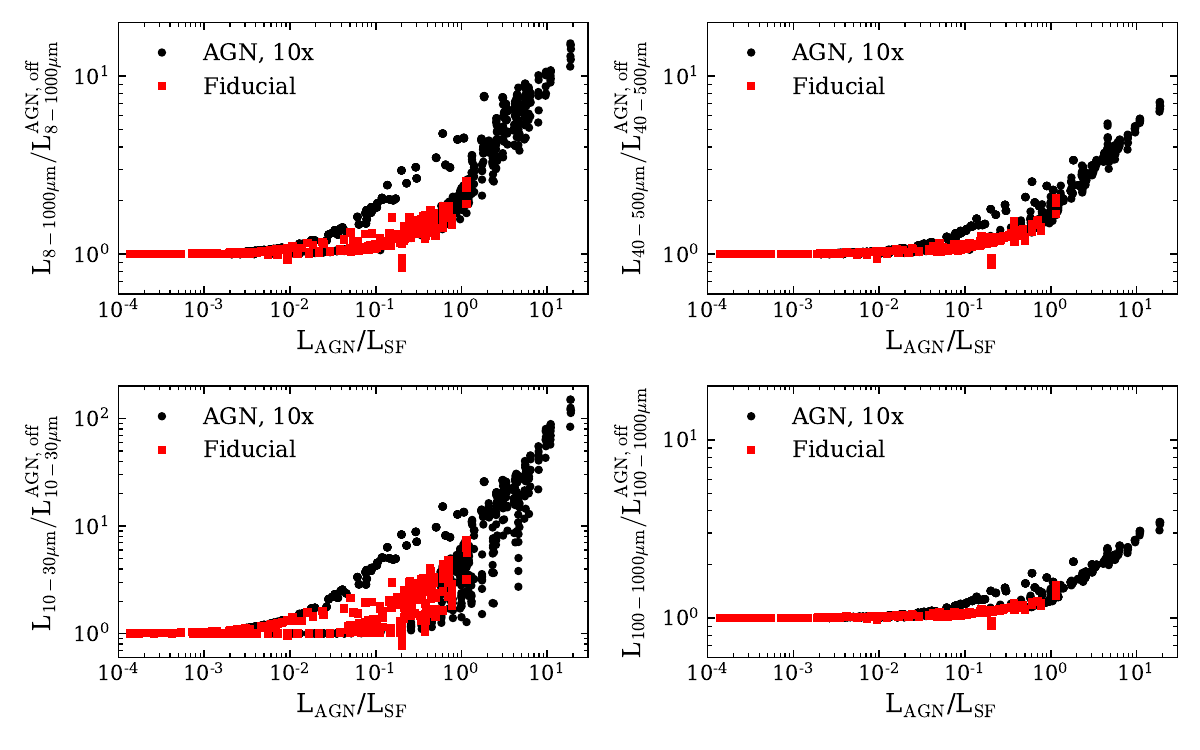}
	\caption{The ratio of the IR luminosity integrated in different bands for \jm{RT calculations with AGN+host galaxy-powered dust emission (`fiducial' or `AGN 10x') relative to those with only host galaxy dust emission (`AGN off')} vs. the ratio of bolometric AGN to star formation luminosity. Data from fiducial (AGN 10x) runs are shown as red squares (black circles). In each panel, we show data for all time snapshots and viewing angles. When the luminosity of the
	AGN is ten times that of star formation, it can boost the total IR (8-1000 \micron) luminosity by an order of magnitude and the FIR (40-500 \micron) luminosity by a factor of $\sim7$. The cold-dust FIR luminosity (100-1000 \micron),
	traditionally assumed to be powered exclusively by star formation, can be boosted by a factor of three. \cchb{Therefore, depending on the precise wavelength range used, the SFR inferred from the IR or FIR luminosity will overestimate the true SFR by a factor of $3-10$.}
	\label{fig:lumRatio}}
\end{figure*}

\begin{figure}
    \gridline{\fig{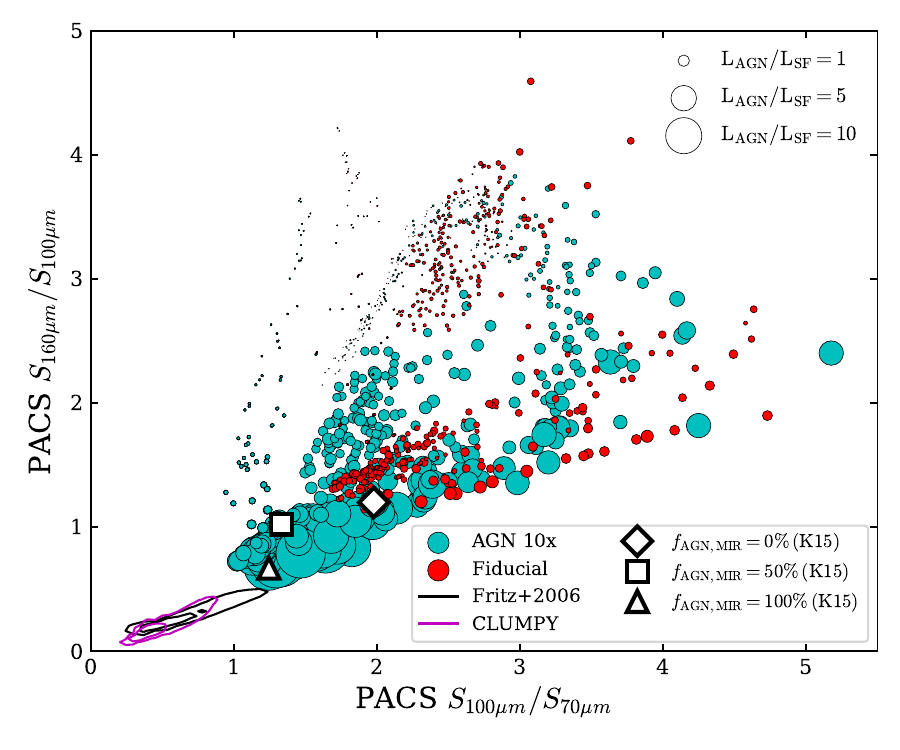}{0.45\textwidth}{}}
    \vspace{-3em}
    \gridline{\fig{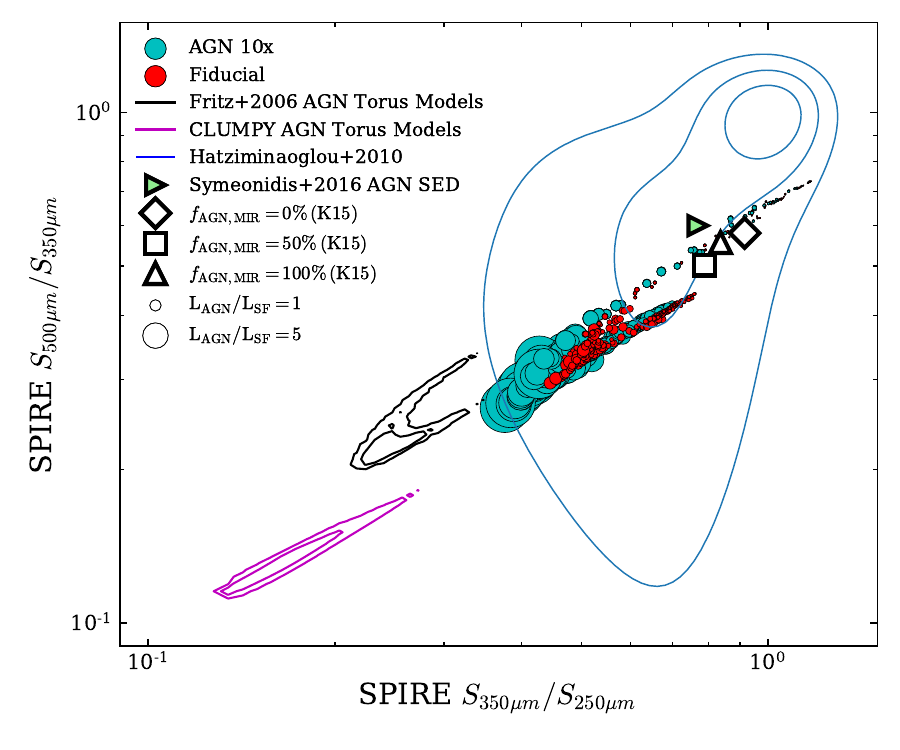}{0.45\textwidth}{}}
    \vspace{-2em}
    \caption{\textit{Top:} \textit{Herschel} PACS 100$\mu$m/70$\mu$m vs. 160$\mu$m/100$\mu$m colors (assuming $z=2$). Cyan and red points correspond to the AGN 10x and fiducial runs, respectively, \jm{and the size of each marker is proportional to \lagnbol/\lsfbol}. For comparison, we show the colors of empirical templates from \cite{Kirkpatrick2015} (K15) at \fagn$=0\%$ (diamond), \fagn$=50\%$ (square), and \fagn$=100\%$ (triangle). Magenta and black contours contain 95\% and 68\% of the distribution in Herschel colors spanned by the \citep{Nenkova2008a,Nenkova2008b} and \cite{Fritz2006} dusty AGN torus models, respectively. \textit{Bottom:} \textit{Herschel} SPIRE
    350$\mu$m/250$\mu$m vs. 500$\mu$m/350$\mu$m colors (assuming $z=2$) following the same color scheme as the upper panel. Blue contours contain the $z\sim2$ AGN-hosting galaxy sample of \cite{Hatziminaoglou2010}. Colors from the \cite{Kirkpatrick2015} SEDs are shown as described in the \textit{Top} panel, and we also include the SPIRE colors of the AGN SED from \cite{Symeonidis2016} (green triangle). The FIR colors of the simulated galaxies are redder (cooler) than those of \emph{all} of the torus models and consistent with those of the observed galaxies, suggesting the `cold' FIR colors of these AGN hosts are not necessarily indicative of ongoing star formation; instead, the FIR emission may be predominantly powered by \cchb{dust-enshrouded} AGN.
    \label{fig:colors}}
\end{figure}

Having demonstrated that the AGN can power cold-dust emission on host-galaxy scales, we now quantify how the AGN affects the IR luminosity in different bands throughout the simulation.
Taking the ratio of \lir\ in the AGN-on and AGN-off simulations captures the fractional change in \lir\ due to dust emission powered by photons of AGN origin. In Figure \ref{fig:lumRatio}, we
plot the ratio of integrated IR luminosities between the AGN-on and AGN-off simulations in four different wavelength bands versus \lagnbol/\lsfbol. The total (far-)IR luminosity from $8-1000\,\mu$m ($40-500\,\mu$m) \jm{is boosted}
by a factor of $\sim10$ ($\sim7$) as \lagnbol/\lsfbol~increases from 1 to $\sim10$. The rise in \lir\ with \lagnbol/\lsfbol\ is mostly driven by increased warm dust emission at shorter wavelengths, as shown in the $10-30\,\mu$m panel of Fig.~\ref{fig:lumRatio}. However, the integrated $100-1000\,\mu$m luminosity can be increased by the AGN by as much as a factor of $\sim3$, \jm{demonstrating that even the coldest dust emission can be powered by the AGN}. This result is at odds with the notion 
that cold dust emission at these FIR wavelengths is predominantly heated by young stars even when an AGN dominates the bolometric luminosity of the galaxy
\citep[e.g.,][]{Stanley2017,Shangguan2020}.

FIR colors are often used to distinguish AGN-powered and stellar-powered dust emission, as IR-selected galaxies that host AGN tend to have warmer IR colors than those that do not (e.g., \citealt{deGrijp1987,Sanders1988};
but cf. \citealt{Younger2009}). For this reason, we present \emph{Herschel} FIR color-color plots of the simulation runs in Figure \ref{fig:colors}, \jm{while also tracking the ratio of AGN luminosity relative to the star-formation luminosity (\lagnbol/\lsfbol). High \lagnbol/\lsfbol\ tends to result in bluer colors; however, this is modulated by changes in extinction with time and viewing angle. Higher levels of extinction ($\mathrm{A_V}$) for fixed \lagnbol/\lsfbol\ yield redder FIR colors.}
For comparison with observations, we also show the empirically derived galaxy SED templates of \cite{Kirkpatrick2015}, which span a range in AGN contribution to the mid-IR emission,
as well as the $z\sim2$ Type 1 and Type 2 AGN sample of \cite{Hatziminaoglou2010}.
The simulated galaxies have FIR colors on the Wien side of the dust distribution (PACS photometry, Fig.~\ref{fig:colors}, \textit{top}) consistent with the observed galaxies. \cchb{The SPIRE colors (Fig.~\ref{fig:colors}, \textit{bottom}), which trace the dust peak, of the simulated galaxies are somewhat \cchb{bluer (warmer)} than those of the majority of the observed galaxies by a factor of $\sim2-3$ when \lagnbol/\lsfbol$\,>1$,
but they are redder (colder) than those of \emph{any} of the torus models in these two widely used torus SED libraries because the torus models do not include
galaxy-scale cold dust emission.
These results suggest that the cool FIR colors of the observed AGN hosts are not necessarily evidence that their FIR cold-dust emission is powered by ongoing star formation;
rather, the AGN itself may be the dominant power source for the cold-dust emission.}

\begin{deluxetable}{lcccc}
    \tablecaption{Maximal boosting of host-galaxy FIR dust emission by a \cchb{dust-enshrouded} AGN at selected wavelengths, and for integrated IR luminosities. \label{tab:tab1}}
    \tabletypesize{\footnotesize}
    \tablehead{$\lambda_{obs}$ & $z=0$ & $z=1$ & $z=2$ & $z=3$}
    \startdata 
    $70\,\mu m$  & 3.8 & 10.8& 30.9 & 47.4 \\[.3ex]
    $100\,\mu m$ & 2.9 & 5.8 & 11.9 & 25.0 \\[.3ex]
    $160\,\mu m$ & 2.4 & 3.4 & 5.2  & 8.5 \\[.3ex]
    $250\,\mu m$ & 2.2 & 2.6 & 3.3  & 4.3 \\[.3ex]
    $350\,\mu m$ & 2.3 & 2.3 & 2.7  & 3.2 \\[.3ex]
    $500\,\mu m$ & 2.6 & 2.2 & 2.4  & 2.6 \\[.3ex]
    \hline
    \hline
     & Fiducial & AGN-10x & & \\[.3ex]
    $\mathrm{L_{IR}(8-1000\mu m)}$ & 2.5 & 14.4 &  &  \\[.3ex] 
     $\mathrm{L_{FIR}(40-500\mu m)}$ & 2.0 & 7.2 &  & \\[.3ex]
     $\mathrm{L(10-30\mu m)}$& 6.7  & 142.7 & &\\[.3ex]
     $\mathrm{L(100-1000\mu m)}$ & 1.5 & 3.2 & & 
    \enddata
    \tablecomments{Maximal boosting as a function of $\lambda_{obs}$ is derived from the redshifted AGN-10x SED considering all timestamps and viewing angles. Generally, these maximum values occur around the most extreme snapshot in our simulations, where \lagnbol/\lsfbol$>10$, and should be interpreted as the maximum possible correction an IR SFR would require to accurately recover the true dust-enshrouded SFR under such admittedly extreme conditions.   }
\end{deluxetable}

\section{Discussion\label{sec:discussion}}

\subsection{Implications}\label{sec:implications}

Our simulations suggest that in extreme cases, AGN may dominate the FIR luminosity of a galaxy at all wavelengths.
The central AGN in our simulations heats dust at radii greater than the half-mass radius of the system, on kpc scales, at times following merger coalescence. Moreover, our simulations span a similar parameter space in FIR colors compared to observations but are redder than all dusty torus models from the CLUMPY \citep{Nenkova2008a,Nenkova2008b} and \cite{Fritz2006} AGN torus SED libraries, instead exhibiting colors that could be interpreted to indicate pure star formation. Consequently, without robust exclusion of a central deeply \cchb{dust-enshrouded} AGN, even cold-dust emission 
at $\ga 100$ \micron~may \cchb{may be `contaminated' by AGN-powered dust emission and thus not yield accurate SFRs}.

Although the current consensus is that star formation dominates cold-dust emission at $\lambda \ga 100 \micron$ (see Section \ref{sec:intro}), some empirical studies have suggested that AGN play an important role in powering dust emission even at these wavelengths \citep[e.g.,][]{Sanders1996}. For example, \cite{Symeonidis2016}, \citet{Symeonidis2017} and \citet{Symeonidis2018} show that optical through sub-mm emission in $z<0.18$ QSOs and $z\sim1-2$ extremely luminous IR galaxies can be dominated by AGN-heated dust out to kpc scales in the host galaxy; however, the degree of contribution by AGN to \lir\ for spatially unresolved systems can depend on assumptions made during SED modeling. Different methods find opposite conclusions based on the shape of the input AGN template SED in the FIR, as demonstrated in \cite{Stanley2017}, who find negligible AGN contribution to sub-mm flux densities for $z\lesssim0.5$ QSOs.
For the simulations presented in this work, \lir\ and $\mathrm{L_{FIR}}$ can be boosted at maximum by factors of $\sim10$ and $\sim7$, respectively, by AGN-powered cold dust emission (Table \ref{tab:tab1}), which lends qualitative support to observational results favoring higher AGN contributions to FIR/sub-mm fluxes. This effect may lead to overestimates of IR-derived SFRs by a factor of $\sim2-3$ on average for the most extreme cases of AGN-heated dust at $z\lesssim3$ in galaxies with high \ledd\ and \lagnbol/\lsfbol$>1$. Spectroscopic SFR diagnostics that do not depend on dust heating (e.g., [Ne\,II], [Ne\,III]) are better suited for measuring accurate SFRs in such cases (e.g., \citealt{Ho2007}); however, these lines can still suffer from significant attenuation in the mid-IR in extremely obscured cases \citep{Smith2007}.

\begin{figure}
    \centering
    \includegraphics[width=0.47\textwidth]{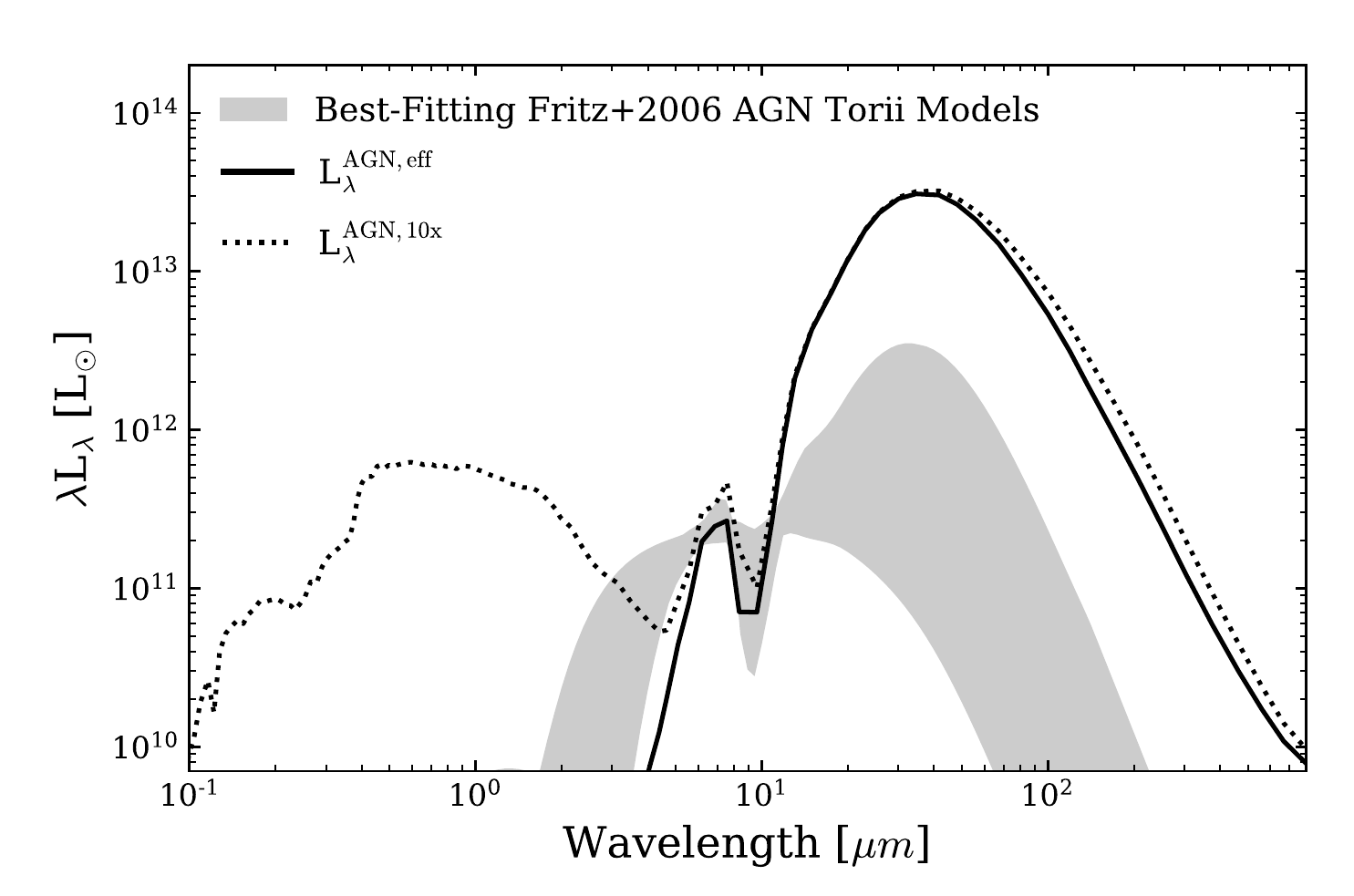}
    \caption{
    The total (\emph{dotted black line}) and effective AGN (\emph{solid black line}) SEDs already shown in Figure \ref{fig:sed}, but now compared against the dusty AGN torus models from \cite{Fritz2006}. We fit the entire model library to the AGN 10x spectrum between $5-13$~\micron\ and show the domain of the top 100 best-fitting models as a shaded grey region. 
    The torus model can reproduce our effective AGN SED's mid-IR dust emission from $\sim2-10$ \micron~but drastically underestimates the FIR emission powered by the AGN.
    Thus, inferring the AGN contribution to the FIR using standard torus models would result in overestimating the SFR.\label{fig:sedFit}
    }
\end{figure}

For the reasons mentioned above, indicators of dust-obscured AGN activity are especially important when \cchb{attempting to measure co-eval} stellar and SMBH mass assembly. X-ray observations can \cchb{identify} Compton-thick QSOs in dusty galaxies out to high redshifts  (e.g., \citealt{Brandt2015} and references therein). \cchb{For example, using \textit{Chandra} observations, \cite{Vito2020} discovered highly obscured AGN residing within two dusty, star-forming galaxies resident in a $z\sim4$ protocluster, one of which is a Compton-thick QSO that has a luminosity comparable to the most-luminous QSOs known. The results presented in the present work suggest that these
obscured AGN may power an appreciable fraction of the cold-dust emission from these galaxies. This example
demonstrates the value of} X-ray follow-up of systems purportedly harboring high levels of obscured star-formation. Alternatively, spectroscopic methods in the mid-IR ($\sim3-30\,\mu$m) may be used to discriminate between \cchb{dust-enshrouded} AGN and star formation using high-excitation emission lines \citep{Spinoglio1992} or polycyclic aromatic hydrocarbon (PAH) features + mid-IR continuum decomposition (e.g., \citealt{Pope2008,Kirkpatrick2015}). Presently, mid-IR spectroscopic techniques are the most sensitive method for identifying \cchb{dust-enshrouded} AGN at all levels of AGN strength relative to the luminosity of the host galaxy \citep{Hickox2018}. Using, amongst other simulations, the fiducial run discussed in this work, \cite{Roebuck2016} find the fraction of $8-1000\,\mu$m luminosity attributed to AGN (the IR AGN fraction) to be a good predictor of the bolometric AGN fraction at times up to coalescence and at $A_V>1$. For such \cchb{dust-obscured} AGN phases, \cchb{quantifying the AGN fraction using mid-IR data  may enable subtracting host-galaxy AGN-heated dust emission to derive accurate SFRs for AGN hosts} (e.g., \citealt{Kirkpatrick2017}). 

Spectroscopic methods and spatially resolved data for more galaxies will improve our understanding of obscured mass assembly; however, a revision of SED-fitting techniques is necessary to fully leverage the power of future observatories. In particular, our results suggest that in highly
dust-obscured galaxies, the host-galaxy-scale
dust emission cannot be considered decoupled from
emission from the AGN, as is done
in community-standard SED fitting codes. As we show in Figure \ref{fig:sedFit}, simpler IR SED decomposition methods also suffer from the lack of AGN-heated cold dust, as it is common to subtract an AGN torus model from the FIR SED of a galaxy and assume the residual power arises from star formation alone. If we make this assumption for our simulation and subtract the best-fitting \cite{Fritz2006} AGN torus model from the total SED, then the residual IR emission would overestimate the true SFR by a factor of $\sim7$ in the extreme case where \lagnbol/\lsfbol\ is maximal. SED fits which include an AGN may also need to consider an associated FIR cold-dust component that does not contribute to the inferred SFR. 

\subsection{Limitations of this work}

We have presented a `case study' of a single galaxy
merger simulation. As emphasized above, the
simulation is representative of $z \sim 2-3$
massive, rapidly star-forming, highly
dust-obscured galaxies (SMGs), and we thus
expect our results to only apply to the bright
end of the IR luminosity function. It would be
valuable to repeat our analysis using simulations
that span a much broader range of properties and
redshifts. 

As discussed in previous works \citep{Younger2009,Hayward2011,Snyder2013}, the treatment of sub-grid dust structure in the ISM is a key uncertainty in RT calculations such as those used here. For the same set of simulations and radiative transfer calculations discussed in our work, \cite{Snyder2013} investigate the effects of varying the prescription for sub-grid dust structure on the time-dependent SED. Notably, they compare the fiducial ISM model adopted in our work, which assumes a uniform dust distribution across each resolution element, with a ``multi-phase'' prescription that assigns some time-dependent fraction of the gas to dense, cold clouds ($\sim27\%$ at $t_{peak}$). In the multi-phase model, attenuation and emission from the (sub-grid) dense cold clouds are ignored (i.e., it is assumed that they have negligible volume filling factor).
They find that in the highly obscured regime, the FIR luminosity is insensitive
to this assumption, and we have confirmed that our results are not qualitatively affected by the treatment of sub-grid dust clumpiness.
Nevertheless, a key avenue for future work is to repeat
this analysis on simulations with much higher
resolution and more sophisticated treatments
of star formation, BH accretion, and stellar
and AGN feedback \citep[e.g.,][]{Angles-Alcazar2020}.
Such an analysis would enable us to check whether our conclusion that
dust-enshrounded AGN can dominate host-galaxy-scale cold-dust emission
holds true for significantly more realistic simulations.


\section{Summary and Conclusions\label{sec:conclusions}}
We have analyzed synthetic SEDs of a hydrodynamical galaxy merger simulation generated via dust RT calculations to investigate the influence heavily \cchb{dust-enshrouded} luminous AGN may have on host-galaxy, kpc-scale cold dust emission.
Our main conclusions are as follows: 
\begin{enumerate}
    \item \cchb{Heavily dust-enshrouded} AGN may power significant cold-dust FIR emission on host-galaxy (kpc) scales; in this particular extreme simulation, the  AGN can boost the total IR luminosity by an order of magnitude and the FIR (100-1000 \micron)
    luminosity by a factor of 3 (i.e., the AGN can be the primary heating source of cold dust).
    \item Our simulations have FIR colors longward of the Wien peak of dust emission consistent with those of $z\sim2$ dusty galaxies, including AGN hosts and `purely star-forming' galaxies.  The FIR colors of our simulations tracing the dust peak are somewhat warmer than observed galaxies when \lagnbol/\lsfbol\ is high.
    However, in both regimes, the FIR colors of the AGN-powered dust emission are redder than those of any of the widely used AGN torus SED models that we considered.
    \item Our results may have significant consequences for IR- and (sub)mm-selected galaxies, which are more likely to feature significant host-galaxy obscuration of the AGN they host than are less-dust-obscured galaxies. For such systems, if a deeply \cchb{dust-enshrouded} AGN is present, the FIR emission may be powered by both young stars and the AGN. Applying standard torus models, which do not include host-galaxy-scale AGN-powered cold-dust emission, to decompose the SED may thus result in overestimating the SFR and underestimating the AGN fraction. 
\end{enumerate}

This work should be considered a `case study' that demonstrates the need for caution when interpreting the FIR SEDs of galaxies that may host \cchb{dust-enshrouded} AGN. Spatially resolved observations (from, e.g., ALMA and \textit{JWST}) and additional AGN diagnostics should be sought to confirm whether a \cchb{dust-enshrouded} AGN is present when studying such systems. Moreover,
there is a need to develop AGN SED models that account for dust obscuration and emission from both the clumpy torus \emph{and} the host galaxy rather
than treating the AGN and its host as decoupled.
\newline 

\footnotesize We thank the anonymous referee, M. Symeonidis, K. Phadke, and E. Daddi for insightful comments on the manuscript, which helped us significantly improve this work. The Flatiron Institute is supported by the Simons Foundation. HAS and JRM-G acknowledge partial support from NASA Grants NNX14AJ61G and NNX15AE56G, and from SOFIA grant NNA17BF53C (08$\_$0069). LR acknowledges the SAO REU program, funded by the National Science Foundation REU and Department of Defense ASSURE programs under NSF Grant AST-1659473, and by the Smithsonian Institution. The simulations in this paper were performed on the Odyssey cluster supported by the FAS Research Computing Group at Harvard University. This research has made use of NASA’s Astrophysics Data System.

\bibliography{references.bib}

\end{document}